\documentclass[journal]{IEEEtran}
\usepackage{cite}
\usepackage{amsmath,amssymb,amsfonts}
\usepackage{algorithmic}
\usepackage{graphicx}
\usepackage{textcomp}
\usepackage{subfig}
\ifCLASSINFOpdf
\else
\fi
\begin{document}
%
\title{Time Division Multiplexing: \\From a Co-Prime Sampling Point of View}
%
%
%

\author{Usham V. Dias
        
\thanks{This work has been submitted to the IEEE for possible publication. Copyright may be transferred without notice, after which this version may no longer be accessible.}%
}

\maketitle

\begin{abstract}
Co-prime sampling is a strategy for acquiring the signal below the Nyquist rate. The prototype and extended co-prime samplers require two low rate sub-samplers. One of the sub-samplers in the extended co-prime scheme is not utilized for every alternate co-prime period. Therefore, this paper proposes a time multiplexing strategy to utilize the vacant slots. It describes the deviation from the existing theory. Closed-form expressions for weight functions are provided. Acquisition of two signals is described with three samplers as well as two samplers. A generalized structure is also proposed with an extremely sparse co-prime sampling strategy.
\end{abstract}


%
\IEEEpeerreviewmaketitle

\section{Introduction}
\label{sec:introduction}
\IEEEPARstart{A}{cquisition} of signals with very large bandwidths require high rate analog-to-digital converters. Several researchers are investigating the use of low rate samplers to achieve results comparable with the high rate samplers. Minimum redundancy array~\cite{1.22} and minimum hole array~\cite{MHA} are some of the works reported from a sub-Nyquist antenna array  perspective. Later, nested and co-prime arrays were proposed~\cite{4.52,4.7} with mathematical expressions to describe its properties. Co-prime sampling is a strategy that can be used to estimate the statistics of a signal at Nyquist rate with sub-Nyquist samplers~\cite{4.7}. The prototype co-prime sampler has two low rate samplers with inter-sample distance $Md$ and $Nd$ where $(M, N)$ is co-prime and $d$ is the Nyquist period. The co-prime sampling strategy has been used for several applications in the literature~\cite{4.7,UVD_PHD,4.45, F2, F3,F5,F1,F4, 16.1,4.40,4.28,4.32,Speec, ShapeCirc1,multidim3,mimo1, mimo2, mimo3,mimo4,doppler1,spacetime1,CR1}. Several modifications to the nested and co-prime array have also been reported~\cite{Nested1,Nested2,Nested3,Nested4,Nested5,Nested6,20.1,4.44,20.4,20.2,20.3,20.5,20.6,20.7,semiCA,CAMPs,UVD_supernyquist,UVD_ExSCA}. Large latency for estimation of the statistics of a signal was a cause of concern. Researchers have also investigated low latency estimation using the co-prime scheme~\cite{I1,S1, UVD_PHD, UVD_CACIS,rect1,UVD_supernyquist,UVD_MUSIC}.

\begin{figure}[!t]
	\centering
	\includegraphics[width=0.5\textwidth]{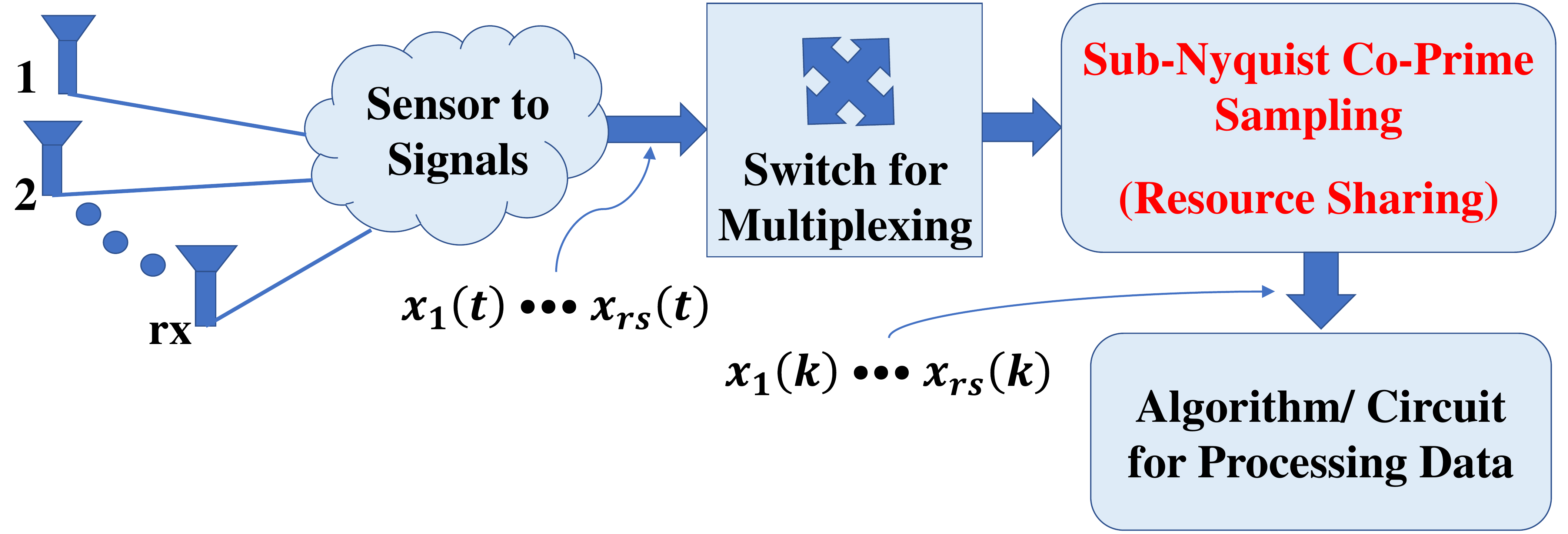}%
	\caption{A broad picture to describe the idea.}
	\label{fig:Abstract}
\end{figure}

This paper analyzes the co-prime sampling strategy for time division multiplexing. One of the early modifications to the prototype co-prime array is the extended co-prime array/sampler proposed in~\cite{4.62}. It is proposed since the prototype co-prime strategy has missing difference values/lag in the co-prime period. The work in~\cite{UVD_extended} develops the extended co-prime theory using the combined signals with low latency. It may be observed in Fig.2-\cite{UVD_extended} that one of the samplers is turned on-off every co-prime period $MNd$. This motivates the idea for time multiplexing of the extended co-prime sampler. An example of time division multiplexing front-end can be found in~\cite{TDM1}. The broad idea is shown in Fig.~\ref{fig:Abstract}. The focus of this paper is to describe time-multiplexing of the co-prime sampler. Next we will briefly describe time division multiplexing for the Nyquist case followed by the extended co-prime sampler and the generalized extremely sparse co-prime sampler.
\section{Time Division Multiplexing}
\begin{figure}[!t]
	\centering
	\subfloat[Independent sampling]{
		\includegraphics[width=0.5\textwidth]{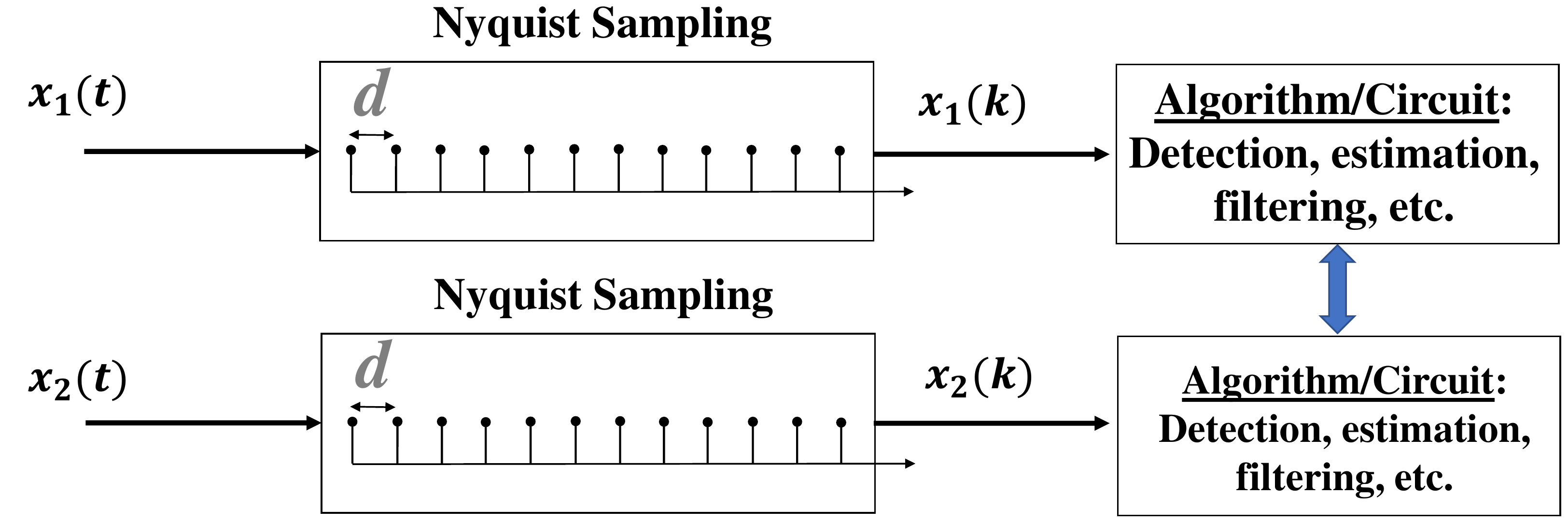}%
		\label{ConceptNyq1}}
	\hfil
	\subfloat[Time multiplexed sampling]{
		\includegraphics[width=0.5\textwidth]{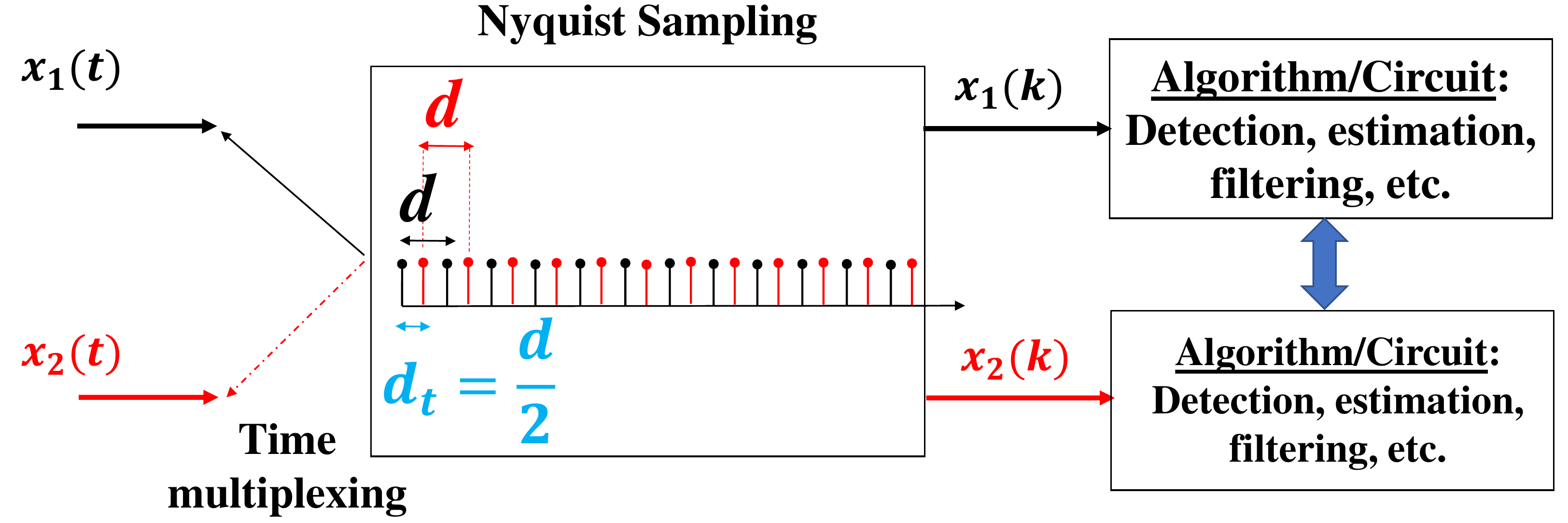}%
		\label{ConceptNyq2}}
	
	\caption{Nyquist sampling of two different signals/random processes $x_1(t)$ and $x_2(t)$.}
	\label{fig:Nyquist}
\end{figure}
\begin{figure*}[!t]
	\centering
	\subfloat[Extended co-prime sampling (Independently)]{
		\includegraphics[width=0.95\textwidth]{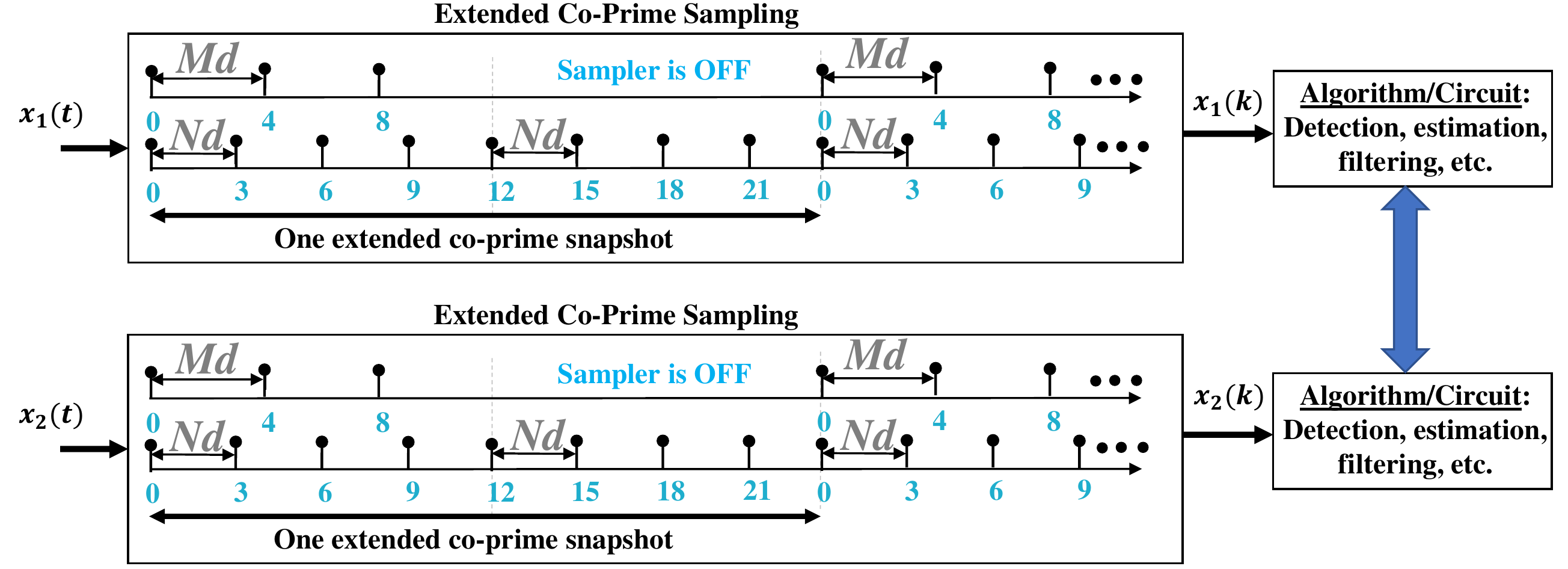}%
		\label{ConceptExtd1}}
	\hfil
	\subfloat[Extended co-prime time multiplexed sampling]{
		\includegraphics[width=0.95\textwidth]{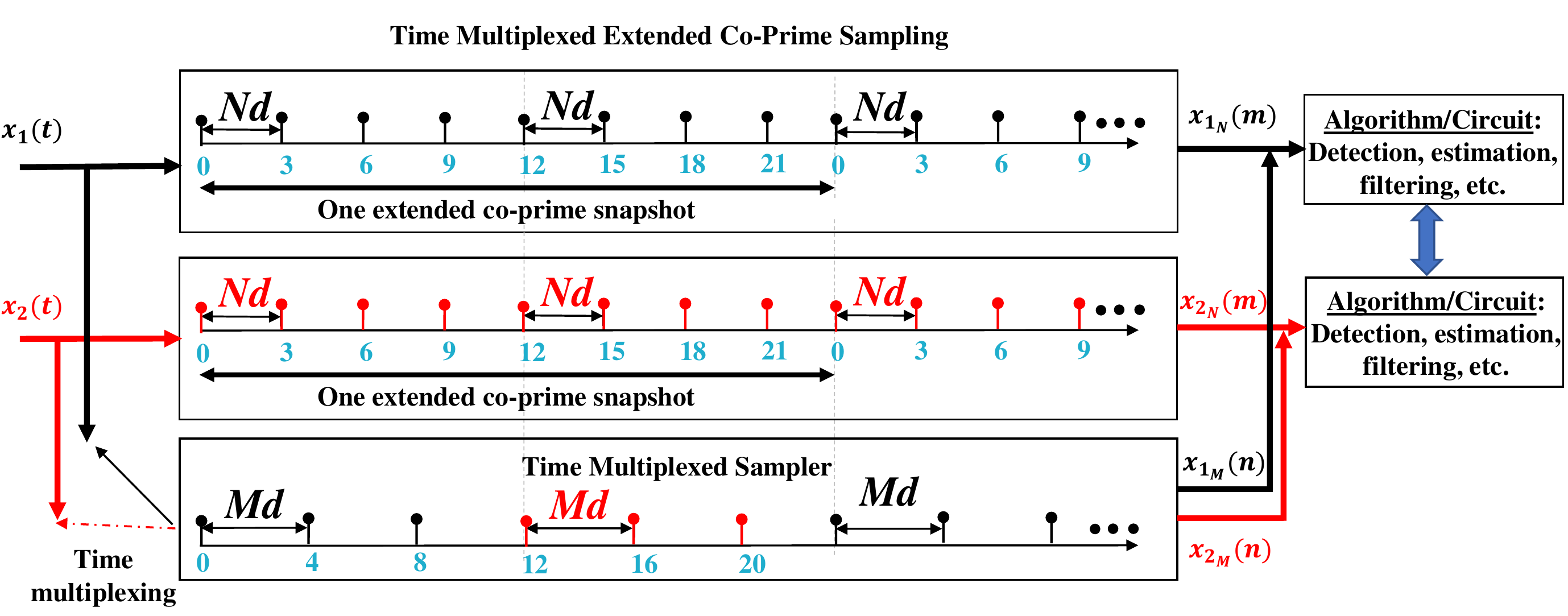}%
		\label{ConceptExtd2}}
	
	\caption{Extended co-prime sampling of two different signals/random processes $x_1(t)$ and $x_2(t)$.}
	\label{fig:Extd12}
\end{figure*}
\begin{figure}[!t]
	\centering
	\subfloat[Self difference set for $x_2(t)$ for sampler with $Md$ spacing]{
		\includegraphics[width=0.4\textwidth]{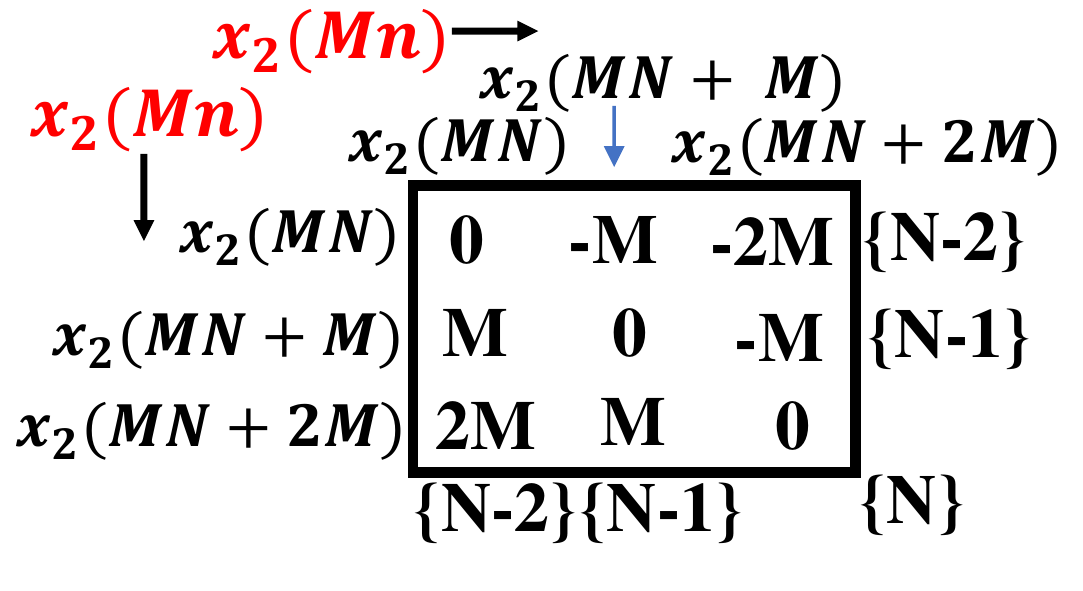}%
		\label{SelfMn_x2}}
	\hfil
	\subfloat[Cross difference set $\mathcal{L}^+_c$ for $x_2(t)$]{
		\includegraphics[width=0.48\textwidth]{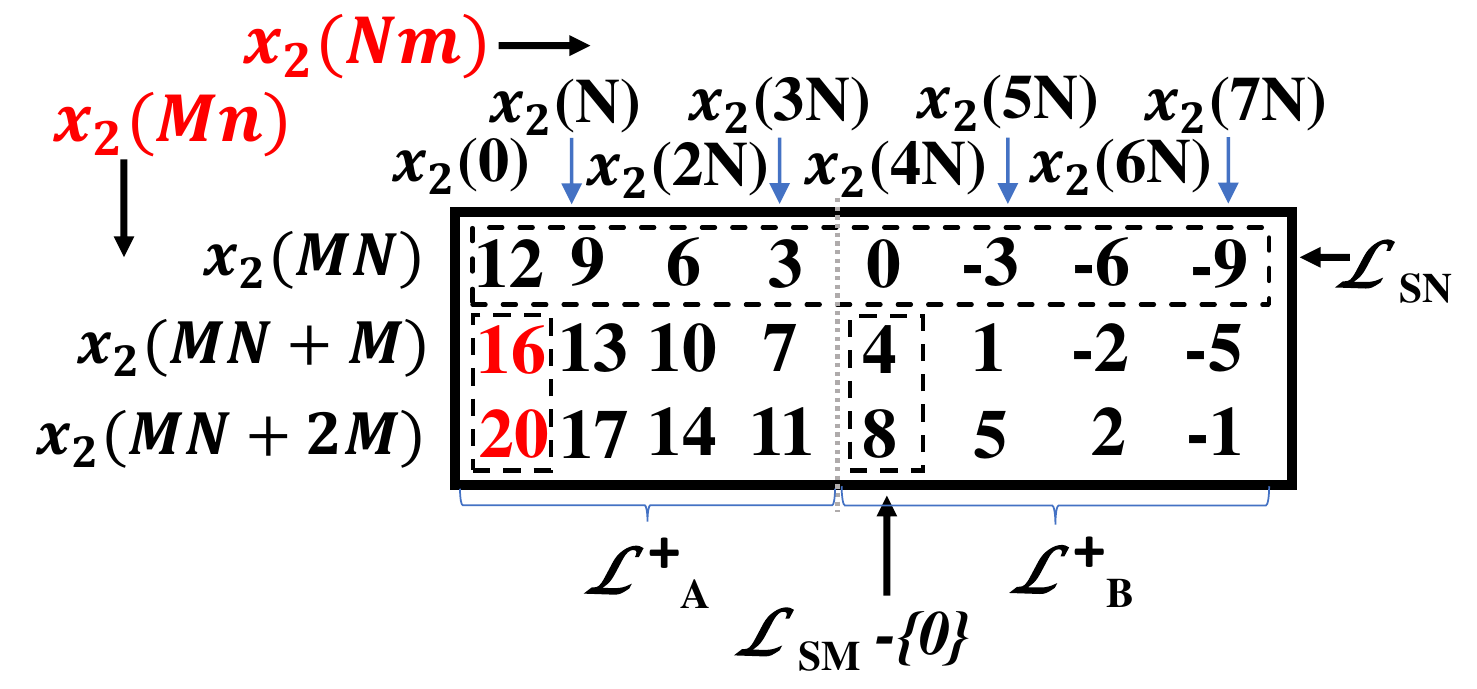}%
		\label{Cross_x2}}
	
	\caption{Extended co-prime sampling of two different signals/random processes $x_1(t)$ and $x_2(t)$.}
	\label{fig:Sets}
\end{figure}
\subsection{Uniform Nyquist Sampling}

The traditional uniform sampling can be employed to acquire multiple signals as shown in Fig.~\ref{ConceptNyq1}. However, a single sampler can be employed with twice the Nyquist rate by time multiplexing the sampler as shown in Fig.~\ref{ConceptNyq2}. It shows two signals $x_1(t)$ and $x_2(t)$. In general multiple signals can be acquired by time multiplexing the sampler. This depends on the availability of a higher rate sampler. $x_1(k)=\{x_1(t) | t=kd\}$ and $x_2(k)=\{x_2(t) | t=kd\}$ represent the acquired samples in Fig.~\ref{ConceptNyq1}. $d$ is the Nyquist sampling period. In Fig.~\ref{ConceptNyq2}, $x_2(k)=\{x_2(t) | t=kd+d_t\}$. The acquired data can be used for detection and estimation of the second order statistics, filtering, etc. Note that certain applications may require the estimation of cross-correlation or cross-spectrum. Therefore, the algorithms may require access to both $x_1(k)$ and $x_2(k)$.
\subsection{Extended Co-Prime Sampling}
The extended co-prime sampling has two sub-samplers with $M$ and $N$ times lower sampling rate. Refer Fig.2 in~\cite{UVD_extended}. Let us assume that there are two different signals or random processes $x_1(t)$ and $x_2(t)$ which need to be sampled. Extended co-prime sampling will require (2+2)=4 sub-samplers. As shown in Fig.~\ref{ConceptExtd1}. Note that one of the sub-samplers in each case is off for one co-prime period $MNd$. This slot can be used to acquire the other signal. Fig.~\ref{ConceptExtd2} demonstrates this time multiplexing idea. Now three samplers are required for implementation. Traditionally, an extended co-prime structure has one sub-sampler with samples at location $Mnd$ with $0\leq n\leq N-1$ i.e. the first co-prime period $MNd$. The other sub-sampler has samples at location $Nmd$ with $0\leq m \leq 2M-1$ i.e. first and second co-prime period $2MNd$. This can be observed in Fig.~\ref{ConceptExtd2} for the signal $x_1(t)$ (marked with black). The theory for this case has been well studied in~\cite{UVD_extended}. Below we mention the expressions for the acquired signal $x_1(t)$:
$x_{1_M}(n)=\{x_1(t) | t=Mnd\}~\text{and}~x_{1_N}(m)=\{x_1(t) | t=Nmd\}$.
The combined signal $x_{1_C}(k)$ at Nyquist rate is given by:
\begin{align}
x_{1_C}(k)&=\left\{
\begin{array}{ll}
x_{1_M}(n), \qquad~\text{for}~ k=Mn, n=[0,N-1]\\
x_{1_N}(m), \qquad~\text{for}~ k=Nm, m=[0,2M-1]\\
0, \qquad \text{otherwise}
\end{array}
\right.      
\end{align}
\begin{figure*}[!t]
	\centering
	\includegraphics[width=0.95\textwidth]{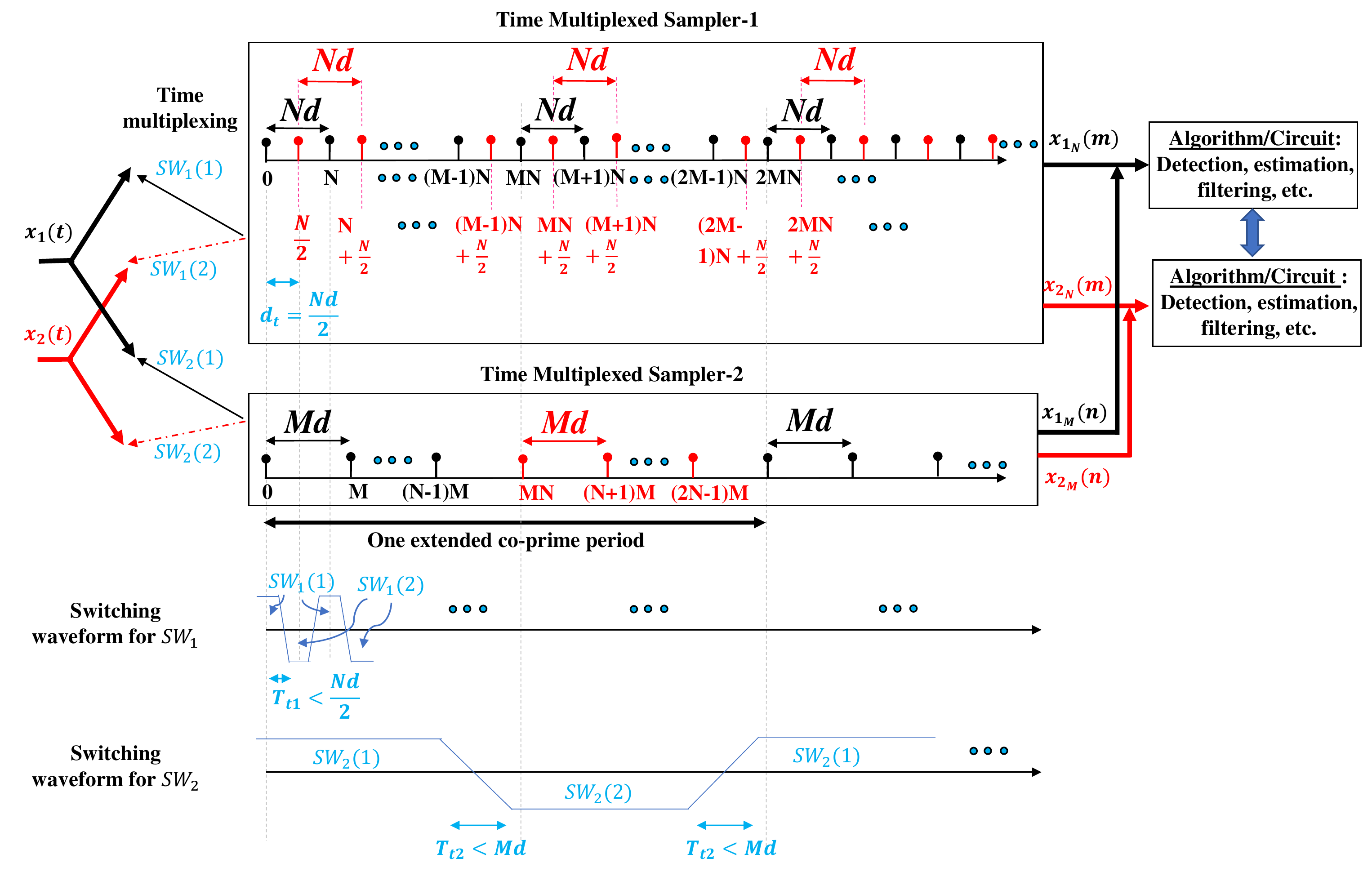}%
	\label{ConceptExtd4}	
	\caption{Time multiplexed extended co-prime acquisition with two samplers.}
	\label{fig:ConceptExtd4}
\end{figure*}
\begin{figure}[!t]
	\centering
	\subfloat[Self difference set for $x_2(t)$ for sampler with $Md$ spacing]{
		\includegraphics[width=0.42\textwidth]{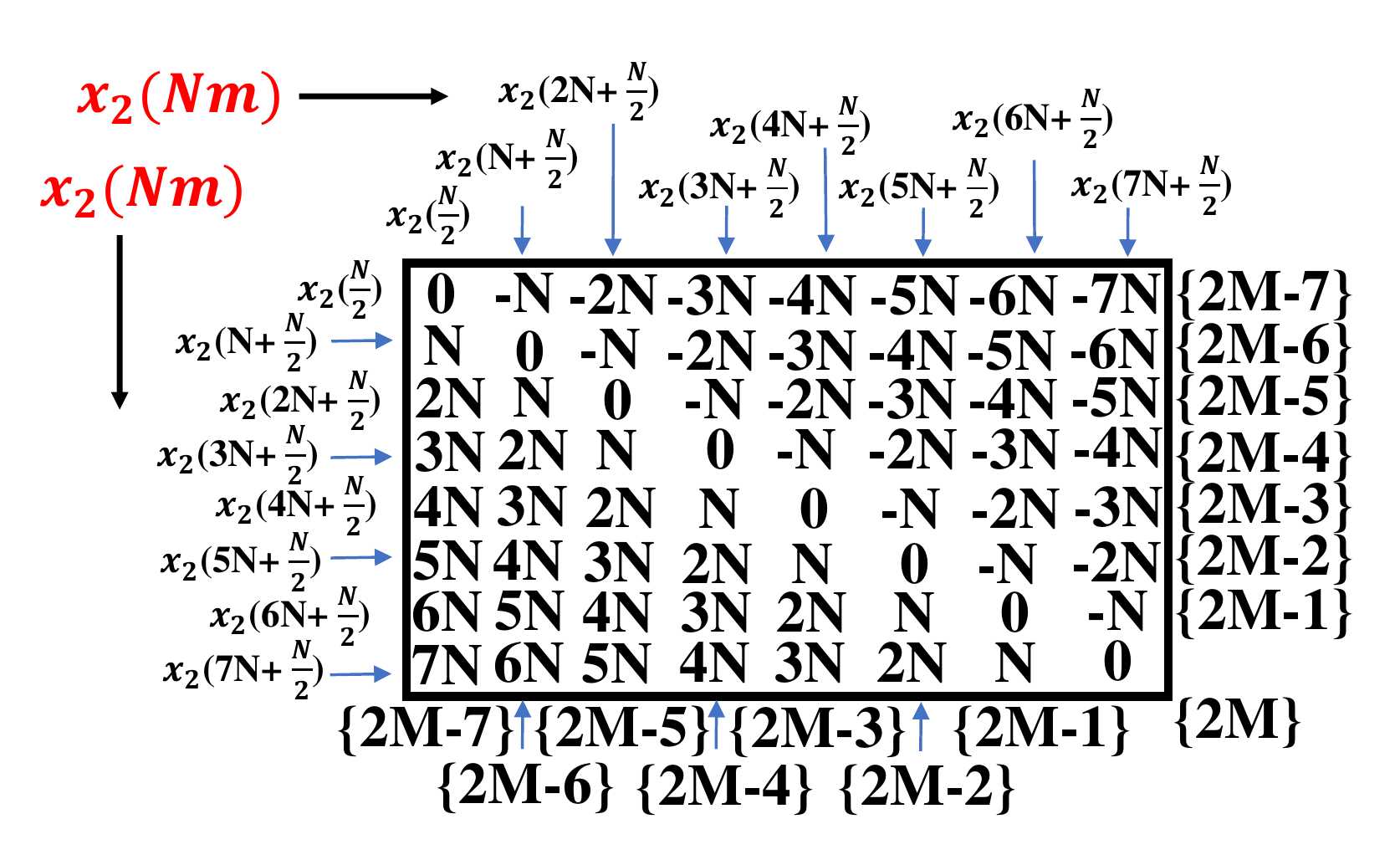}%
		\label{SelfMn_x2_super}}
	\hfil
	\subfloat[Cross difference set $\mathcal{L}^+_c$ for $x_2(t)$]{
		\includegraphics[width=0.4\textwidth]{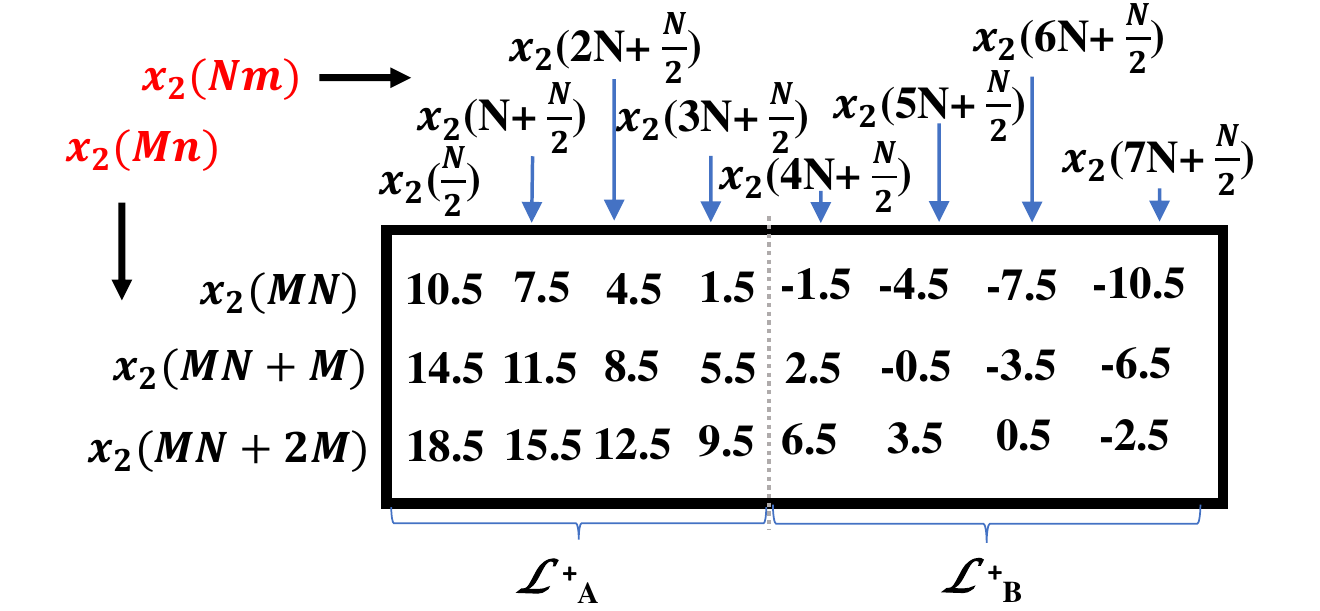}%
		\label{Cross_x2_super}}
	
	\caption{Extended co-prime sampling difference sets for $x_2(t)$ for the two sampler scenario.}
	\label{fig:Sets_super}
\end{figure}

Now observe the signal $x_2(t)$ (marked in red) in Fig.~\ref{ConceptExtd2}. One sub-sampler acquires samples for two periods $Nmd$ with $0\leq m \leq 2M-1$. The other samples are at locations $Mnd$ but with $n$ in the range $N\leq n \leq 2N-1$. $x_{2_M}(n)=\{x_2(t) | t=Mnd\}~\text{and}~x_{2_N}(m)=\{x_2(t) | t=Nmd\}$. Keep in mind that the range of $n$ is different when compared with $x_1(t)$. The combined signal $x_{2_C}(k)$ at Nyquist rate is given by:
\begin{align}
x_{2_C}(k)&=\left\{
\begin{array}{ll}
x_{2_M}(n), \qquad~\text{for}~ k=Mn, n=[N,2N-1]\\
x_{2_N}(m), \qquad~\text{for}~ k=Nm, m=[0,2M-1]\\
0, \qquad \text{otherwise}
\end{array}
\right.      
\end{align} The question that needs to be answered is whether the theory developed in~\cite{UVD_extended} would be valid here. To this end it is important to analyze the difference sets and weight function/number of contributors for estimation.

Let us consider an example with $M=4$, $N=3$. In addition, we can assume $d=1$ throughout the paper without affecting the behaviour for general values of $d$. For $x_1(t)$, the sub-sampler with inter-element spacing $Nd$ has samples at locations $[0, 3,6,9,12,15,18,21]$ and sub-sampler with $Md$ spacing has samples at $[0, 4,8]$. Refer Fig.~\ref{ConceptExtd2} (black samples). The self difference and cross difference matrix is given in Fig.3 and Fig.4 in~\cite{UVD_extended} respectively. The weight function and bias window is given by equation (1) and (10) in~\cite{UVD_extended} respectively. Now for $x_2(t)$ (Fig.~\ref{ConceptExtd2} marked in red), the sub-sampler with inter-element spacing $Nd$ has samples at locations $[0,3,6,9,12,15,18,21]$. This is same as the locations for $x_1(t)$ hence the self difference matrix is same. The second sub-sampler with spacing $Md$ has samples at $[12,16,20]$. The self difference matrix is given in Fig.~\ref{SelfMn_x2}. Note that the contents of the matrix here is same as (Fig.3a in~\cite{UVD_extended}). The cross difference matrix $\mathcal{L}^+_c$ is given in Fig.~\ref{Cross_x2}. The area marked in red is a deviation from the previous known theory. Let us compute the number of pairs for autocorrelation estimation $z_1(l)$ and $z_2(l)$ (for $x_1(t)$ and $x_2(t)$ respectively). This can be computed as described in Fig.4.6 in~\cite{UVD_PHD}. The combined sampling points/pattern for $x_1(t)$ is:
$p_1(k)$=[1,0,0,1,1,0,1,0,1,1,0,0,1,0,0,1,0,0,1,0,0,1,0,0]
and $x_2(t)$ is:
$p_2(k)$=[1,0,0,1,0,0,1,0,0,1,0,0,1,0,0,1,1,0,1,0,1,1,0,0].
Therefore, $z_1(l)$=[10  2  2  7  3  2  6  1  2  5  1  1  4  1  1  3  0  1  2  0  0  1  0  0] and $z_2(l)$=[10  2  2  7  2  2  6  1  1  5  1  1  4  1  1  3  1  1  2  0  1  1  0  0] for $l=[0, 2MN-1]$. Note that $z_1(l)$ and $z_2(l)$ is symmetric for negative lags i.e. $l=[-1, -2MN-1]$. $z_1(l)\neq z_2(l)$ for $l=[\pm4 \pm8 \pm16 \pm20]$ i.e. for $l=\pm Mn$ with $n=[1,2,\cdots N-1, N+1,N+2,\cdots2N-1]$. In fact, $z_1(l)=z_2(l)+1$ for $l=\pm Mn$ with $n=[1,2,\cdots N-1]$ and $z_1(l)=z_2(l)-1$ for $l=\pm Mn$ with $n=[N+1,N+2,\cdots2N-1]$. $\sum z_1(l)=\sum z_2(l)=100$ for the example considered. $z_2(l)$ has 2$(N-1)$ additional unique lag values, i.e. $\pm16$, $\pm20$ for the example considered. The expression for the number of contributors $z_1(l)$ is given by (1) in~\cite{UVD_extended}. While the expression for $z_2(l)$ is given below:
\begin{equation}\label{eq:extend_entirex2}
\begin{split}
z_{2}(l)=\underbrace{\sum\limits_{n=-(N-1)}^{N-1} (N-\mid n\mid)\delta(l-Mn)+\sum\limits_{n=N+1}^{2N-1}\delta(\mid l \mid-Mn)}_\text{A}\\
+\underbrace{\sum\limits_{m=-(2M-1)}^{2M-1} (2M-\mid m\mid)\delta(l-Nm)}_\text{B}\\
+\underbrace{\sum\limits_{n=N+1}^{2N-1}\sum\limits_{m=M+1}^{2M-1} 2 \delta(l-(Mn-Nm))-\delta(l)}_\text{C}\\
+\underbrace{\sum\limits_{n=N+1}^{2N-1}\sum\limits_{m=1}^{M-1} \delta(\mid l\mid-|Mn-Nm|)}_\text{D}
\end{split}
\end{equation}
It is straight forward to obtain the correlogram bias window in this case as the Fourier transform of $z_2(l)$.

Next let us implement the time multiplexing for extended co-prime sampling with two samplers. This is shown in Fig.~\ref{fig:ConceptExtd4}. For $x_1(t)$ (marked in black), both the samplers have the same sampling locations as described in Fig.~\ref{ConceptExtd2}. Therefore the self and cross differences are same. For $x_2(t)$ (marked in red) in Fig.~\ref{fig:ConceptExtd4}, the sampler associated with switch $SW_2$ has same sampling locations as Fig.~\ref{ConceptExtd2}. Therefore the self difference is same. But the sampler associated with switch $SW_1$ has a different sampling location.  $x_{2_M}(n)=\{x_2(t) | t=Mnd, n=[N, 2N-1]\}~\text{and}~x_{2_N}(m)=\{x_2(t) | t=Nmd+\frac{Nd}{2}, m=[0,2M-1]\}$. As an example, let $M=4$ and $N=3$. The sampling locations are $[1.5,4.5,7.5,10.5,13.5,16.5,19.5,22.5]$. The self difference matrix is given in Fig.~\ref{SelfMn_x2_super}. Note that this matrix is same as that in (Fig.3b in \cite{UVD_extended}) and has integer values (multiples of $N$). However, the cross difference matrix does not have integer values as shown in Fig.~\ref{Cross_x2_super}. This relates to the concept of super Nyquist sampling~\cite{UVD_supernyquist}. This implies that $x_2(t)$ has some samples at multiples of $\frac{d}{2}$. Hence, the bandwidth of $x_1(t)$ and $x_2(t)$ need not be the same. Fig.~\ref{fig:ConceptExtd4} also provides the switching waveform. The transition time $T_{t1}$ and $T_{t2}$ depends on $Nd$ and $Md$. The hold time at the location $SW_{1}(1)$, $SW_{1}(2)$, $SW_{2}(1)$, and $SW_{2}(2)$ also depends on $N$, $M$, and $d$.

Next let us discuss a more generalized set-up for sampling known as the Extremely Sparse Co-Prime Arrays/Samplers (ExSCA)~\cite{UVD_ExSCA}.
\begin{figure*}[!t]
	\centering
	\subfloat[ExSCA sampling (Independently)]{
		\includegraphics[width=0.95\textwidth]{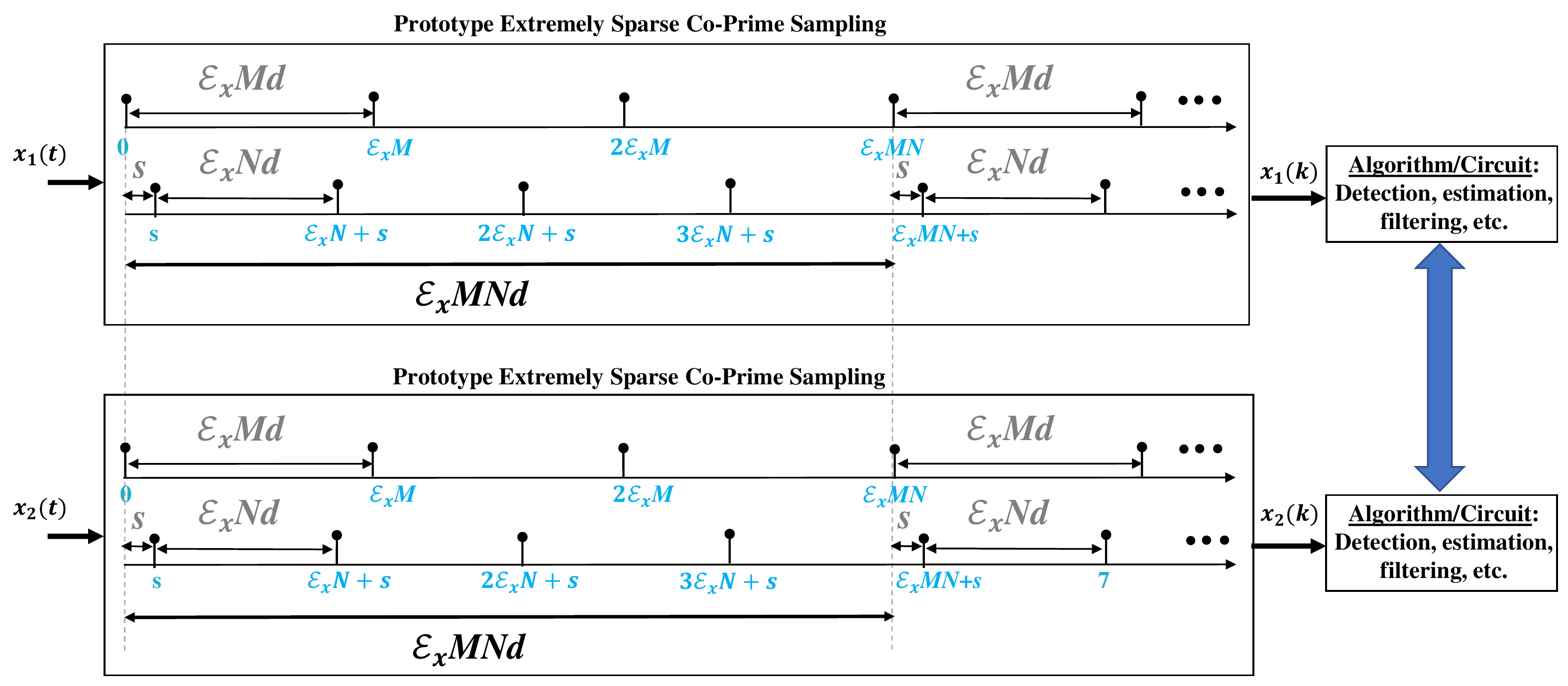}%
		\label{ConceptExSCA1}}
	\hfil
	\subfloat[ExSCA time multiplexed sampling]{
		\includegraphics[width=0.95\textwidth]{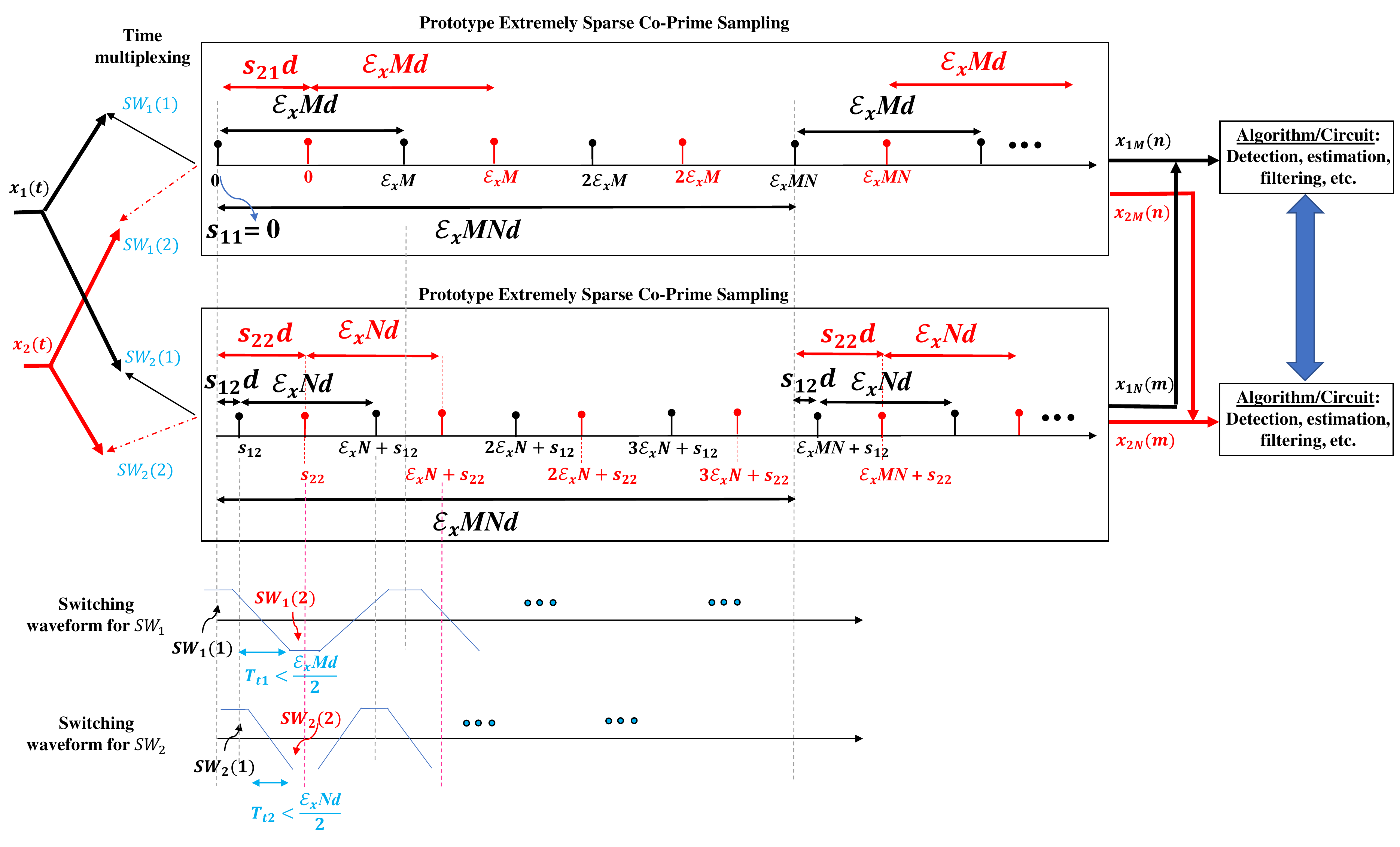}%
		\label{ConceptExSCA2}}
	
	\caption{ExSCA-based sampling of two different signals/random processes $x_1(t)$ and $x_2(t)$.}
	\label{fig:ExSCA12}
\end{figure*}

\begin{figure}[!t]
	\centering
	\includegraphics[width=0.5\textwidth]{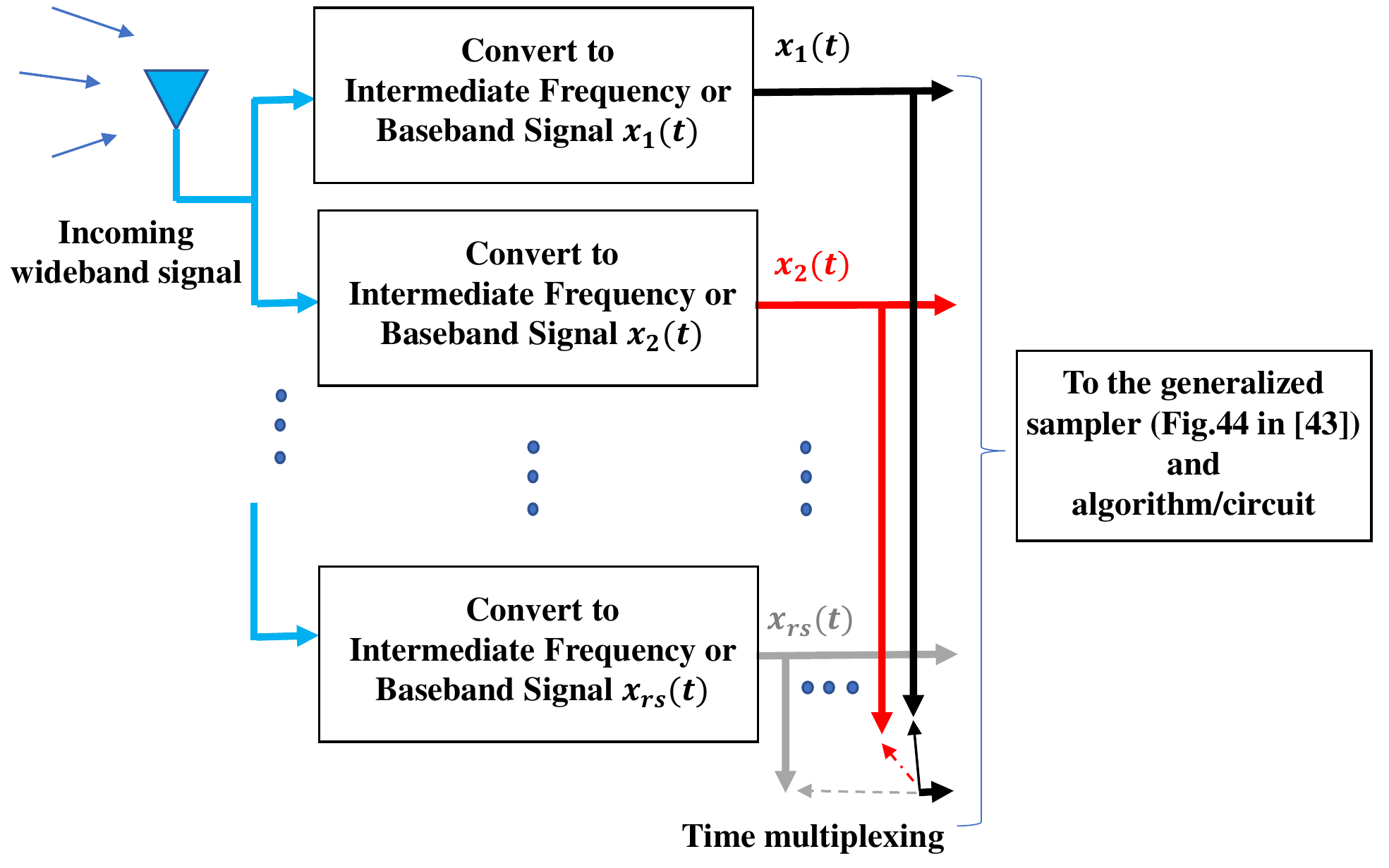}%
	\label{ConceptExtd3}	
	\caption{Single antenna wideband signal acquisition and statistics estimation of narrow band signals/random processes.}
	\label{fig:ConceptExtd3}
\end{figure}
\begin{figure}[!t]
	\centering
	\includegraphics[width=0.5\textwidth]{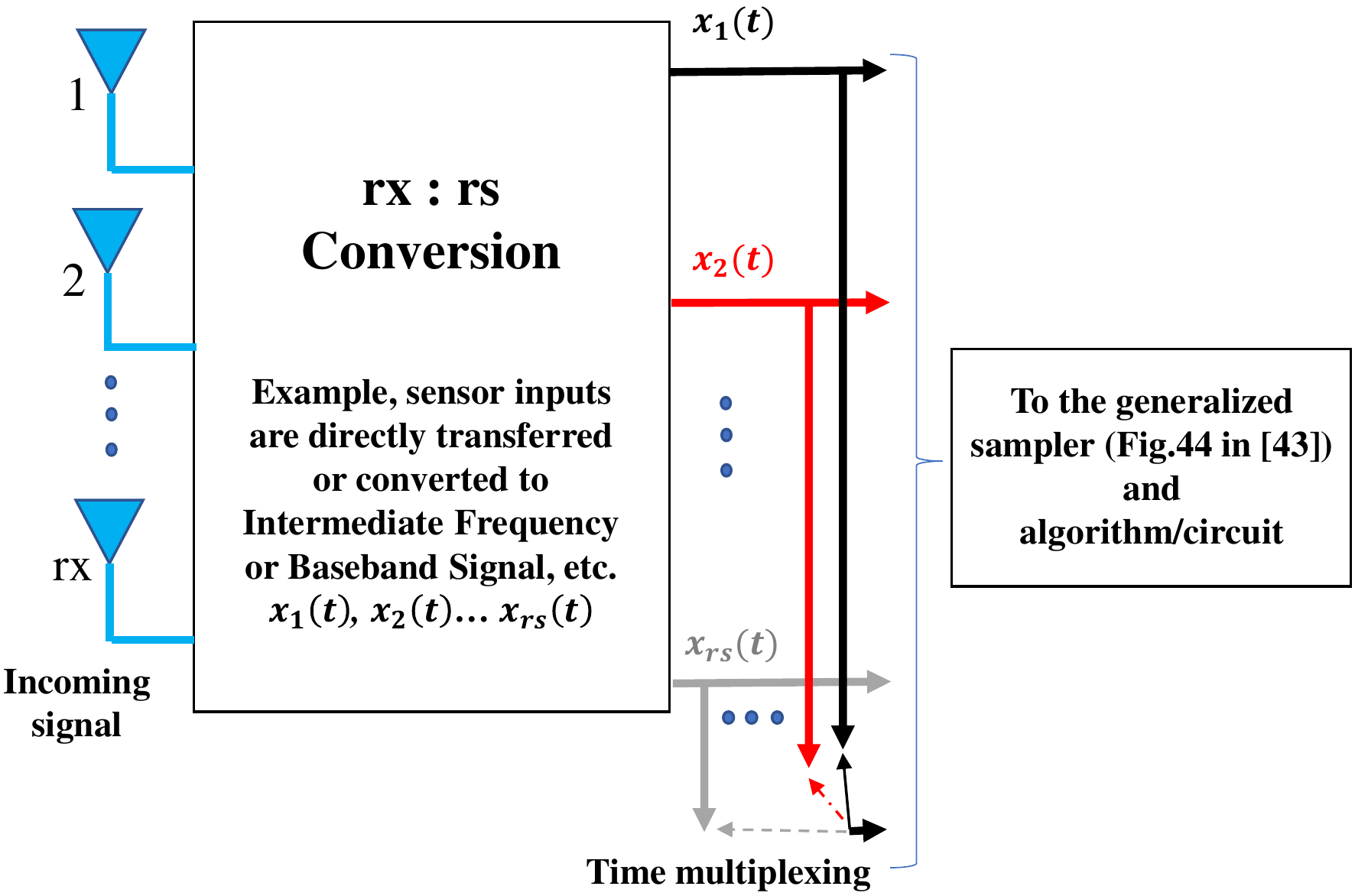}%
	\label{ConceptGeneralized}	
	\caption{Generalized acquisition from $rx$ sensors and processing of $rs$ signals/random processes.}
	\label{fig:ConceptGeneralized}
\end{figure}

\section{Generalized Acquisition and Sampling Strategy}
In the simplest form, the ExSCA sampling strategy uses factors $\mathcal{E}_xM$ and $\mathcal{E}_xN$ as inter-sample distance. Two different signals can be acquired using the ExSCA strategy as in Fig.~\ref{ConceptExSCA1}. It requires four samplers. However, if a co-prime pair of samplers $(M, N)$ is available, then ExSCA can be easily implemented with two samplers as shown in Fig.~\ref{ConceptExSCA2}. The switching waveforms are provided. It is important to note that one of the locations of $x_2(t)$ is coinciding in the two samplers. This is bad since it can lead to aliasing as described in~\cite{UVD_ExSCA}. Therefore, the parameters need to be selected wisely.  
For the example in Fig.~\ref{ConceptExSCA2}, $s_{ij}$ represents the shift of sampler-j from the origin for signal $x_i(t)$. Normally, $s_{11}$ can be assumed as the origin, i.e. $s_{11}=0$. The shift for the second signal $x_2(t)$ with sampler-1 is $s_{21}=s_{11}+\frac{\mathcal{E}_xM}{2}$. For the second sampler, $s_{12}$ needs to be designed to prevent aliasing and maximize information (along with $M$, $N$, and $\mathcal{E}_x$). $s_{22}=s_{12}+\frac{\mathcal{E}_xN}{2}$. As an example, if $\mathcal{E}_x=2$, $M=4$, $N=3$, $s_{11}=0$, and $s_{12}=1$, then $s_{21}=4$ and $s_{22}=4$. However, these parameters are not a good choice since the samples for $x_2(t)$ overlap. The example here has two signals $x_1(t)$, $x_2(t)$ and two samplers. However, three signals can also be sampled with just two co-prime samplers using the Extremely sparse strategy with $\mathcal{E}_x=3$.

It may also be noted that $\mathcal{E}_xd$ can be designed to have different Nyquist sampling rate for $x_1(t)$ and $x_2(t)$. $d$ represents the smallest Nyquist period corresponding to the signal with the highest bandwidth. Here, super Nyquist sampling may be viewed as a special case of ExSCA. In a more generalized setting, we can have $q$ samplers with different factors $\mathcal{E}_1$, $\mathcal{E}_2$, $\cdots$ $\mathcal{E}_q$. Refer Fig.44 in~\cite{UVD_ExSCA}. Also note that, $x_1(t)$ and $x_2(t)$ are different signals. In general, there could be $rs$ such signals: $x_1(t)$, $x_2(t)$,$\cdots$ $x_{rs}(t)$. However, they may be acquired from a single sensor/antenna as shown in Fig.~\ref{fig:ConceptExtd3}. The hardware can convert the signal to the appropriate intermediate frequency or baseband. Each one of them could have a different bandwidth and hence a different Nyquist sampling period. This can be further time-multiplexed with $q$ different samplers. The $rs$ signals need not be fully connected to all $q$ samplers. 

Furthermore, $r$x sensors/antennas can be used to sense the signal as shown in Fig.~\ref{fig:ConceptGeneralized}. Some of the sensors may perform wideband to multiple narrowband signal conversion while others may not. This would depend on the application.

\section{Conclusion}
This paper describes time division multiplexing for acquisition of signals using the extended co-prime scheme. The difference set is analyzed and closed-form expression for the contributors for estimation is provided. It also describes the implementation of a time multiplexed extremely sparse co-prime sampler using the standard co-prime sub-samplers. The work presented here can sample signals with different bandwidths or Nyquist sampling periods. A single antenna or sensor can be employed to capture the wideband signal which can be converted to the desired narrow-band signals for sampling. It also provides a generalized structure with multiple sensors/antennas which can be employed based on the application. The number of signals to be acquired can be more than, less than, or equal to the number of samplers. This would depend on the constraints of the application.
%




\ifCLASSOPTIONcaptionsoff
  \newpage
\fi



%
%

\bibliographystyle{IEEEtran}
\bibliography{refs}

%

%






\end{document}